# Lissajous scanning magnetic particle imaging as a multifunctional platform for magnetic hyperthermia therapy

James Wells[1*], Shailey Twamley[2,3], Aparna Sekar[2], Antje Ludwig[2,3,4], Hendrik Paysen[1], Olaf Kosch[1], Frank Wiekhorst[1]

[1]Physikalisch-Technische Bundesanstalt, Abbestraße 2-12, 10587 Berlin, Germany

[2] Charité - Universitätsmedizin Berlin, corporate member of Freie Universität Berlin, Humboldt-Universität zu Berlin, Berlin Institute of Health, Medizinische Klinik für Kardiologie und Angiologie, Campus Mitte, Berlin, Germany

[3] DZHK (German Centre for Cardiovascular Research), partner site Berlin, Germany

[4] Charité - Universitätsmedizin Berlin, corporate member of Freie Universität Berlin, Humboldt-Universität zu Berlin, Berlin Institute of Health, Klinik für Radiologie, Berlin, Germany

*Corresponding author: james.wells@ptb.de

## Abstract

The use of engineered nanoscale magnetic materials in healthcare and biomedical technologies is rapidly growing. Two examples which have recently attracted significant attention are magnetic particle imaging (MPI) for biological monitoring, and magnetic field hyperthermia (MFH) for cancer therapy. Here for the first time, the capability of a Lissajous scanning MPI device to act as a standalone platform to support the application of MFH cancer treatment is presented. The platform is shown to offer functionalities for nanoparticle localization, focused hyperthermia therapy application, and non-invasive tissue thermometry in one device. Combined, these capabilities have the potential to significantly enhance the accuracy, effectiveness and safety of MFH therapy. Measurements of nanoparticle hyperthermia during protracted exposure to the MPI scanner's 3D imaging field sequence revealed spatially focused heating, with a maximum that is significantly enhanced compared with a simple 1-dimensional sinusoidal excitation. The observed spatial heating behavior is qualitatively described based on a phenomenological model considering torques exerted in the Brownian regime. In-vitro cell studies using a human acute monocytic leukemia cell line (THP-1) demonstrated strong suppression of both structural integrity and metabolic activity within 24 h following a 40 min MFH treatment actuated within the Lissajous MPI scanner. Furthermore, reconstructed MPI images of the nanoparticles distributed among the cells, and the temperature-sensitivity of the MPI imaging signal obtained during treatment are demonstrated. In summary, combined Lissajous MPI and MFH technologies are presented; demonstrating for the first time their potential for cancer treatment with maximum effectiveness, and minimal collateral damage to surrounding tissues.

## Introduction

Magnetic particle imaging (MPI) is an emerging biomedical imaging technique which maps the spatial distribution and local conditions of magnetic nanoparticle (MNP) tracers [1]. A growing number of commercial and self-built scanners are in operation globally, using a variety of scanning methods and image reconstruction techniques [2] [3] [4] [5] [6] [7]. All MPI variants require a dynamic magnetization measurement of the MNP contained within the sample. The superposition of oscillating excitation fields with static gradient fields allows a spatially resolved signal to be collected for image reconstruction. Single-frame acquisition times are typically short (<<1 s), making the technique attractive for measuring dynamic processes such as blood circulation. Longer acquisition times of (>10 minutes) have also been reported for continuous process monitoring, the averaging of weak signals [8], in systems employing exceptionally low-frequency magnetic fields or in systems employing raster scanning image acquisition. MPI is under evaluation for applications including the mapping of circulation [9], internal bleeding [10], cell tracking [11] and oncology imaging [12]. So far, little attention has been paid to the inevitable energy dissipation which occurs during MPI image acquisition, or the potential impact of MPI imaging sequences on living tissues. For MPI technology to progress past regulators and into the clinic, the limits of safe operation should be ascertained.

The concept of MNP heating when exposed to fast alternating magnetic fields (AMF) was first introduced in the 1950s [13]. Magnetic field hyperthermia (MFH) therapy harnesses this effect as the basis of a novel cancer treatment [14] [15]. First, MNP are accumulated within the tumor tissue, then a high-frequency AMF is applied to the affected region. This excitation causes the MNP to dissipate heat, raising the temperature in the local vicinity. The goal of MFH is a precision therapy which inflicts thermal damage to the tumor cells, while minimizing collateral damage to surrounding healthy tissues. MFH therapy typically aims to achieve temperatures in the range of 42-45 ˚C [16]. In recent decades MFH has gained increasing credence as a succession of encouraging preclinical and clinical trials have been reported [17] [18].

Both MPI and MFH technologies require a combination of magnetic nanoparticles and AMF. However, the magnetic fields implemented typically differ due to the different goals in each case. The goal during MPI is normally to avoid inflicting damage by MNP heating. MPI scanners therefore typically employ frequencies between 1-150 kHz, with the most common scanner at present (Bruker 20/25 FF) operating at frequencies close to 25 kHz. So far, MFH research has largely focused on the heating produced by single-axis, sinusoidal magnetic field excitations at high frequencies. In contrast to this, MPI scanners are typically capable of supplying more complex magnetic field sequences. For 3D imaging, the 20/25 FF MPI scanner produces AC excitations in the x- y- and z- axes which are combined with a static gradient field containing a single field-free-point (FFP). These superimposed fields drive the FFP throughout the imaging volume on a defined trajectory (Lissajous figure). A small body of work has demonstrated the potential for "focusing" MFH therapy using static gradient fields superimposed on simple 1D excitations [19]. The heating induced by a 2D circularly polarized magnetic field has also been briefly studied [20]. However, the heating resulting from the more complex scheme of multifrequency 2D or 3D excitations combined with additional field gradients for Lissajous scanning has not yet been published.

Little research has so far been reported on heating effects within MPI scanners. Exploratory studies using an MPI device based on a raster scanning trajectory have examined the capability for this type of MPI scanner to support the application of MFH therapy [8] [21]. However, these studies focused on using a separate imaging apparatus (single-axis excitation $f$ = 20 kHz) in combination with a second, high frequency (*f* = 350 kHz) excitation for MFH application. In this case the heating produced by the scanner's imaging sequences was shown to be negligible over the course of a 2.5 minute low-frequency imaging scan, while significant heating was demonstrated during the subsequent high-frequency MFH exposure. In addition to the excitation frequency and amplitude employed, the scanning trajectory employed during MPI (Lissajous, raster or other) may also significantly influence the extent to which imaging induces hyperthermia effects.

So far, the heat dissipated by the 3D Lissajous scanning sequences employed in the 20/25 FF MPI scanners has not been studied. In this work, for the first time, the capabilities of Lissajous scanning MPI as a tool for supporting the implementation of MFH are assessed. The heating behavior of MNP samples was measured under different magnetic field sequences and at different locations within the MPI scanner, and the results compared with a simple model based on the magnetic field sequences in each case. The effect of scanner-induced MFH on a human acute monocytic leukemia cell line (THP-1) was then investigated.

# Materials and Methods
## MPI Scanner

A commercial 20/25 FF MPI scanner (Bruker BioSpin, Germany) installed at the Charité university hospital in Berlin was used for all measurements. The AC drive fields have frequencies of 24510, 26042 and 25253 Hz, in the x- y- and z- axes respectively; with variable amplitudes between 0 - 12 mT. A static gradient field for spatial encoding can also be applied. This gradient is variable up to maximum values of 1.25/1.25/2.5 T/m in the x-, y- and z- axes (1:1:2 ratio is always maintained). The gradient field contains a FFP at the center of the scanner's field-of-view (FOV). To achieve 3D imaging, all three excitation axes as well as the field gradient must be used simultaneously. This magnetic field combination results in the FFP traversing a Lissajous trajectory around the FOV, with a period of 21.5 ms. For all MPI work in this study, AC components in each axis were set to either 0 mT or 12 mT. The number of excitation axes switched-on, and the state of the static gradient field, are reported for each measurement using the following notation style: (2D, 2.5 T/m). This example represents x- and y- axes with 12 mT AC excitations switched-on, no AC excitation in the z- axis, and a 2.5 T/m static field gradient. For optimal imaging resolution, maximum values should be used for dynamic and static fields (3D, 2.5 T/m), resulting in a FOV of 2 x 2 x 1 cm$^3$ volume. The MPI scanner is contained within a large shielded room to reduce electromagnetic noise during measurements.

Image reconstruction in the 20/25 FF scanner requires the acquisition of a sizeable calibration measurement (System Function (SF)) for the tracer used [2]. The acquisition of multiple SF for the same tracer under different conditions opens the opportunity for "multi-color" MPI, in which the local temperature [22], binding state [23] or other parameter of the MNP (as differentiated by the SF) can also be resolved. At present, SF measurements at different temperatures have not been implemented in the Berlin scanner. Image reconstructions in this study all used a single SF obtained at room temperature. Using the same SF to reconstruct images of an object measured at different temperatures results in slight deviations in the reconstructed image. The changes in the reconstructed images caused by different sample temperatures were visualized by subtracting the values for each voxel in the image acquired at baseline temperature from those in subsequent images. By summing voxel values along the z-axis, a convenient 2D map or "difference image" is produced, this can be used to gauge changes in the reconstructed image due to temperature. With suitable calibration, this contrast can be translated into temperature resolved MPI images in the future.

To measure MNP heating within the MPI scanner, a sample was mounted on the scanner's testbed surrounded by a foam block for positioning and thermal insulation (See Fig. 1(a)). A fiber-optic thermometer was placed within the sample for temperature monitoring. The thermometer read-out box and other electronics were located outside of the scanner's shielded room. After allowing the sample to reach thermal equilibrium within the scanner, calorimetry measurements were conducted by running a consecutive set of measurements with a total field exposure time of 9 minutes. A powerful cooling system ensures that the temperature within the scanner is kept stable, even during continuous operation of the imaging fields. In the absence of any MNPs, the temperature within the sample space was found to be stable (+/- 0.2 K) during protracted measurements of more than 2 hours.

## Nanoparticles

Ferucarbotran (FCT) (Meito Sangyo, Japan) is a liquid suspension of nanoparticles consisting of dextran-coated iron oxide. The majority of them have hydrodynamic sizes between 45 nm and 60 nm, while they exhibit a bimodal core distribution with peaks at 5.5 nm and 24 nm [24] [25]. FCT was selected for this work as it is a common tracer used within MPI studies, is available commercially in a stable and reproducible form, and because it has been extensively characterized within the literature. FCT is known to dissipate heat under appropriate alternating magnetic field conditions [26]. Evidence for macroscopic and microscopic heating of FCT at frequencies well below those normally used in therapeutic hyperthermia have also been previously reported [27]. Temperature dependence of the MPI imaging signal produced by FCT nanoparticles under simple 1D field excitations was previously noted in [28].

## Specific absorption rate and corrected slope analysis

Characterization of MNP heating efficiencies, particularly via calorimetry measurements, is a common topic in the MFH literature. The measurement result is typically expressed using a quantity commonly referred to as the specific absorption rate (SAR) or specific loss power (SLP) [29]. The SAR is given by the total heating power ($P$) dissipated per unit mass of MNP material ($m_{MNP}$) contained.

$$SAR = \frac{P}{m_{MNP}} \qquad (1)$$

Various methods for calculating the specific absorption rate (SAR) from calorimetry heating curves have been presented in the literature. Here we employ the corrected slope technique (CST) described by Wildeboer et al [29]. The method is designed to account for the thermal losses from the sample during heating. At lower temperatures, heat losses are expected to scale linearly with increasingly elevated sample temperatures, with deviations from linear behavior occurring at higher temperatures. The CST first requires identification of the temperature range in which the linear-loss regime persists. The linear loss regime is dependent not only upon the particular calorimetry apparatus used, but also upon the specific sample volume, shape and material used. Thus, it is necessary to characterize the extent of the linear loss regime for each measurement apparatus and sample size. Once the linear loss range has been identified, calorimetry measurements intended for SAR analysis should be only be conducted within this temperature range. The CST can then be used to find the specific absorption rate from the sample's heating curve via

$$SAR_{CST} = \frac{\left(C\frac{dT}{dt} + L\Delta T\right)}{m_{MNP}} \qquad (2)$$

Where $C$ = specific heat capacity, ΔT =elevation of sample temperature above ambient at the point in heating curve used for SAR calculation, dT/dt = gradient of heating curve at that point in the heating curve, L = linear loss parameter specific to the particular sample and apparatus combination. A fitting tool for estimating L for each measurement curve, and calculating $SAR_{CST}$ is available from Resonant Circuits Ltd, London UK [30]. In contrast to some other SAR estimation methods in the literature (for example, the initial slope method), the corrected slope method also allows multiple SAR calculations to be made along the length of the heating curve, increasing the level of certainty in the result.

## Cell Studies

### Cell cultivation

THP-1 cells (human acute monocytic leukemia cell line) obtained from the ATCC (Wesel, Germany) were cultured in suspension in a humidified incubator at 37 °C with a 5 % $CO_2$ concentration in RPMI medium 1640 (Invitrogen, Karlsruhe, Germany). Culture medium was supplemented with 10 % fetal calf serum (FCS, Biochrom, Berlin, Germany), 100 U/mL penicillin, 100 μg/mL streptomycin (p/s, Invitrogen), and 2 mM L-glutamine (Invitrogen). Exact cell numbers were determined with a hemocytometer.

### Sample preparation and treatment

MNP-loaded loose tumor phantoms were created shortly before MPI treatment. For each, $6·10^6$ THP-1 cells were pelleted and suspended in RPMI medium supplemented with 1 % FCS. FCT was then added to produce an iron concentration of $c$(Fe)= 300 mM in a volume of 400 μl. Two phantom samples were produced for each measurement, one for MPI treatment, and an identical one for control. For MPI treatment of cells, tubes containing warm water circulated from a pump located outside the scanner's shielded room were used to maintain the sample and foam block at a baseline temperature of 37 ˚C within the scanner (Fig. 4 (a)). Phantoms were mounted on the foam block within the MPI scanner and allowed to reach a stable temperature of 37 °C before being exposed to a continuous MPI imaging excitation (3D 12 mT, 2.5 T/m) for 40 min. During this time, the control samples were incubated at 37 °C on a heating block situated outside of the MPI scanner's magnetically shielded room. Following MPI treatment, both samples were washed once with 10 mL PBS and pelleted. Cells were then suspended in 6 mL of RPMI medium supplemented with 10 % FCS and 1 % p/s and cultured in a humidified incubator at 37 °C. Aliquots of cells were analyzed at 1 h ($t_1$), 2 h ($t_2$), 4 h ($t_4$) and 24 h ($t_{24}$).

### Cellular viability measurements

THP-1 viability analysis was performed by flow cytometry (CyAN-ADP flow cytometer, Beckman Coulter). Aliquots of cells were centrifuged and resuspended in PBS containing 0.5 μg/mL DAPI. DAPI positive cells were defined as having lost membrane integrity. After adding 0.05 % Triton X-100 for complete cell membrane permeation the DNA content of the nuclei was measured by DAPI emission. At least 10,000 cells were analyzed for each sample.

The MTT (3-(4,5-dimethylthiazol-2-yl)-2,5-diphenyl tetrazolium bromide) assay was used to estimate the number of viable cells. Metabolically active cells convert MTT to dark blue, water-insoluble MTT formazan and the formation of formazan was used as a measure of viable cells. Aliquots of cells seeded in 96- well culture plates ($10^5$ cells/200 μl RPMI/well) were incubated for 24 h in a cell incubator. After 24 h 10 μL MTT solution from the stock (5mg/mL) was added and cells were incubated in the cell incubator for 1 h. Plates were centrifuged, washed with PBS and the formazan was dissolved by adding 150 μL of isopropanol. The absorption of the solution was measured at 570 nm using a spectrophotometer (Spectramax Molecular Devices).

## Modelling of MPI scanner fields and torques

To explore the mechanism underpinning the experimentally observed SARs, the scanner's magnetic fields were simulated for different excitation sequences. Using the assumption that the AC components produced by the x-, y- and z- axis excitation coils are homogeneous throughout the sample space, the dynamic magnetic field component (D) in each axis at a time (t) is given by

$$D_x = A_x \sin(2\pi f_x t)$$
$$D_y = A_y \sin(2\pi f_y t)$$
$$D_z = A_z \sin(2\pi f_z t) \qquad (3)$$

Where *A* denotes the amplitude, and *f* the frequency of the *x*, *y* and *z* components of the scanner's excitation. In the absence of a magnetic field gradient, these components are sufficient to describe the fields within the scanner.

For each axis, the components of the static gradient field ($S_x$, $S_y$, $S_z$) were generated for equally separated positions in space (*r*) within a 4 cm x 4 cm x 4 cm volume located over the center of the scanner's field of view. For measurements with the gradient field switched off, the remanent field was assumed to be zero in each axis. By adding the static and dynamic components at each location, the total amplitude of the scanner's magnetic field in each axis can be calculated

$$B_x(t,r) = D_x(t) + S_x(r)$$
$$B_y(t,r) = D_y(t) + S_y(r)$$
$$B_z(t,r) = D_z(t) + S_z(r) \qquad (4)$$

These components combine to give the total magnetic field vector $B_T(t,r)$ produced by a given excitation sequence at each location and point in time. The resultant field magnitude for each point in space at time *t* is then given by:

$$|B_T(t,r)| = \sqrt{B_x(t,r)^2 + B_y(t,r)^2 + B_z(t,r)^2} \qquad (5)$$

$B_T(t)$ was calculated at equally spaced time points ($\Delta t_i$ = 1 μs) throughout the magnetic field cycle. For a given location within the scanner (*r*), the average magnetic field amplitude $|\overline{B_T}|$ produced by the coil system during a given excitation sequence can then be found via

$$|\overline{B_T(r)}| = \frac{1}{n}\sum_{i=1}^{n}|B_T(t,r)|_i \qquad (6)$$

Where *n* = number of time points for which $|B_T(t,r)|$ was calculated within the time period.

When the magnetization of a nanoparticle is not aligned with the resultant magnetic field vector, a torque will be exerted on the nanoparticle. For 2D and 3D excitations, the Lissajous path of the magnetic field vector over time means that the angles between magnetic field vectors produced at adjacent time points $B_{T1}$ and $B_{T2}$ varies over the Lissajous cycle. The angle between the vectors for each pair of timepoints at a given point in space is given by:

$$\theta(r) = \cos^{-1}\left(\frac{(B_{T1} \cdot B_{T2})}{|B_{T1}| \cdot |B_{T2}|}\right) \tag{7}$$

Using the calculated values for $|B_T|$ at each point in space and time, and the angular change $\theta$ between adjacent time-point pairs, an approximation of the variations in torque exerted on nanoparticles present within the scanner can be made throughout the magnetic field sequence. To do this, the following assumptions are employed; that Brownian relaxation processes are dominant in the system, that the nanoparticles rotate fast enough in the liquid that by each new timepoint the particle's magnetization has aligned with the previous timepoint (MNP lags behind the field vector by 1 timepoint), that the particle's magnetization ($m$) is constant at all times, and that the relaxation time of the nanoparticles is independent of the spatially varying static gradient field. Based on these assumptions, an estimate of the pattern of varying torque ($\tau$) exerted on a nanoparticle over time can be made for adjacent timepoints throughout the Lissajous cycle using

$$\tau = m \times B \tag{8}$$

The average magnitude of $\tau$ across all the timesteps within a single Lissajous cycle $\overline{|\tau|}$ is calculated for different types of magnetic field sequence. For field sequences which include a static field gradient, $\overline{|\tau|}$ will be spatially dependent.

For nanoparticle systems in which Brownian relaxation dominates, a higher average torque value ($\bar{\tau}$) exerted on MNP inside the scanner is assumed to correspond to a higher level of MNP rotation, and thus a greater overall heat dissipation. To explore the validity of employing the assumptions necessary for the torque-based model for a given system, the Néel ($R_N$), Brownian ($R_B$) and effective ($R_{\text{eff}}$) relaxation times[1] were calculated using

$$R_N = R_0 \exp\left[\frac{KV_c}{k_B T}\right] \tag{9}$$

$$R_B = \frac{3 V_h \eta}{k_B T} \tag{10}$$

$$R_{\text{eff}} = \frac{R_N R_B}{R_N + R_B} \tag{11}$$

where $R_0$ = material specific time constant (typically $10^{-9}$ -$10^{-12}$ s), K= magnetic anisotropy, $V_c$ = magnetic core volume, $k_B$ = Boltzmann constant, $T$ = temperature, $V_h$ = hydrodynamic volume, and $\eta$ = viscosity of carrier liquid.

# Results
*Sequence-dependent hyperthermia at center of Lissajous MPI scanner*
Calorimetry measurements of a 400 µL sample of FCT c(Fe)= 935 mmol/L) placed at the center of the MPI scanner's FOV are shown in Figure 1 for different excitation fields. The sample mounting including temperature probe is shown in Fig. 1(a). Examples of the FCT heating curves produced by a simple (1D, 0 T/m) excitation, and a full Lissajous MPI imaging sequence (3D, 2.5 T/m) are shown in Figure 1(b). A control measurement of a 400 µL pure water sample under (3D, 2.5 T/m) is also shown in Fig. 1(b). The temperature of the water sample remained stable over the full measurement time, showing that the scanner electronics and coils do not leak heat into the sample, and that the heating effects observed in the FCT samples arise purely from MFH effects. For a simple single-axis excitation with no field gradient (1D, 0 T/m), a total temperature rise of 11.7 K was recorded over the 9-minute field exposure. By contrast, the three-axis excitation with maximum field gradient (3D, 2.5 T/m), produced a significantly larger rate of heat dissipation ($\Delta T$ = 20.4 K over 9 minutes). The 3D Lissajous MPI sequences used for high-resolution imaging produce significantly more heat than an equivalent length of exposure to a simple 1D excitation as is normally employed for MFH actuation.

---

[1] Note, in this work $R_i$ instead of the common $\tau_i$ is used to denote relaxation time variables.

To better understand the effect of the multi-axis fields and field gradients, further heating curves were measured for the 400 µL FCT sample under different field sequences within the scanner. A SAR value was calculated for each measurement using the CST. In Fig. 1(c), the SAR values calculated for the (1D, 0T/m), (2D, 0T/m), (3D, 0T/m), (3D, 1T/m), (3D, 2.5T/m) field arrangements are presented. The average value for the linear-loss parameter for 400 µL measurements was 4.8 mW/K. The results show that the addition of an extra dimension of AC excitation, (1D to 2D or 2D to 3D) in the absence of a gradient field is associated with a 19 % increase in the SAR. Further increases in the SAR are observed as the gradient field is introduced in addition to the 3D excitation. The SAR reaches a maximum in the field arrangement (3D, 2.5 T/m), which is the sequence used to provide the best achievable spatial resolution for 3D imaging within this design of preclinical scanner. The results indicate that the SAR at the center of the scanner's FOV during MPI imaging is 60% higher than would be experienced from a simple 1D excitation in the absence of a gradient field.

In Figure 1D, the SAR calculated from heating curves obtained for 1D and 3D excitations both in the absence of a static field gradient, and in the presence of a 2.5 T/m field gradient are compared. The addition of a static gradient field in the presence of a 1D excitation field is seen to have a very different effect on the SAR as compared with when the same static gradient field is superimposed over a 3D excitation. In the case of the 1D excitation, addition of the gradient field reduces the SAR by 46 %. For the 3D excitation the same gradient field increases the SAR by 13 %.

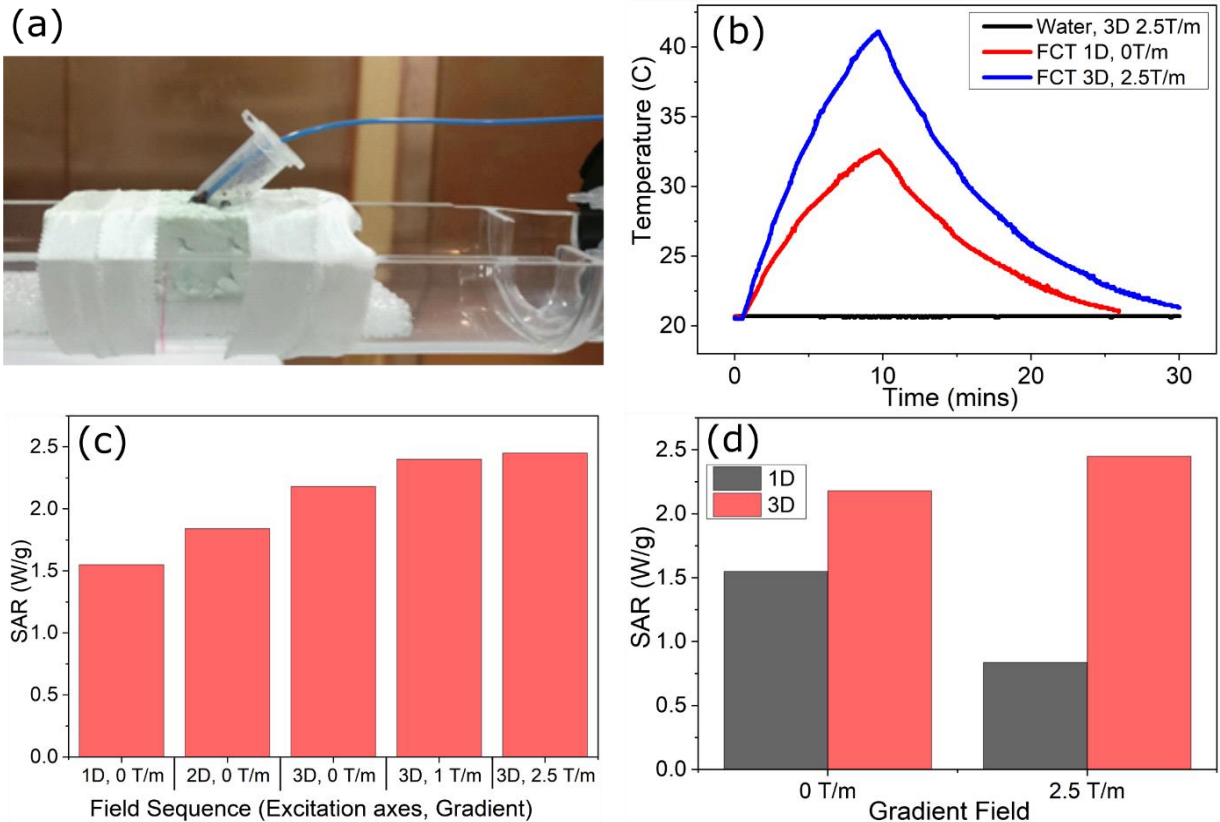

*Figure 1 (a) Photograph of 400 µL FCT sample with fiber-optic temperature probe mounted on scanner bed, ready for placement at the center of the MPI scanner's FOV. (b) Raw heating curves obtained using water (black curve: (3D, 2.5 T/m)) and FCT samples in the scanner under two different excitation sequences (red curve: (1D, 0 T/m) and blue curve: (3D, 2.5 T/m)). (c) SAR values calculated using the corrected slope technique (eq. 2) for the FCT sample under different field arrangements in the scanner. (d) contrast between SAR values in 1D and 3D excitations, and the impact in each case of adding a 2.5 T/m gradient field.*

*Spatial dependence of scanner-induced hyperthermia*

The heating behavior of a smaller sample volume of 50 µL FCT (935 mmol/L) under different magnetic field sequences is presented in Figure 2. Heating effects in this sample were less pronounced, and more erratic, due to greater thermal losses as compared with the 400 µL sample. This is reflected in the larger average linear-loss parameter for 50 µL of 14.4 mW/K. However, the smaller sample size allowed the spatial dependence of the SAR to be probed. The sample was first placed in the center of the scanner's FOV, and heating curves were again measured under different 1D, 2D and 3D excitations (Fig. 2(a-c)). Despite the smaller sample, measurable heating was still observed for all excitation types. The impact on the heating behavior of static field gradients in addition to the AC components was also measured. The SAR was calculated for the 1D, 2D and 3D cases both in the absence of a field gradient, and with maximum field gradient. The calculated SAR values are presented in Fig. 2(d), with arrows displaying the percentage increases and decreases between the different excitation fields. For the 1D excitation, the superposition of a static field gradient is again associated with a reduction in heat dissipation. For the 2D and 3D excitations, the addition of a static gradient field results in an increase in the amount of heat dissipated.

The spatial dependence of the SAR was then investigated at varying positions along the x axis of the MPI scanner. Heating curves were measured at different positions under different magnetic field excitations. The SAR for each of these measurements was determined and the results are presented in figures 2(d) and (e). Measurements with 0 T/m and 2.5 T/m gradient fields were conducted consecutively at each position before moving to the next. In the 1D case, the addition of the gradient field produces a significant focusing effect, with the heating dropping to zero just outside the scanner's FOV. The maximum of the focused SAR is less than the SAR in the absence of a gradient field. For the 3D excitation, the same enhancement of the SAR is observed with the addition of a gradient field. The focusing effect of the gradient field is less strong in this case, with measurable heating occurring up to 4 cm from the center of the FOV. It is not possible to conduct the same measurements in the *y* and *x* axes as the bore of the scanner limits the movement of the sample.

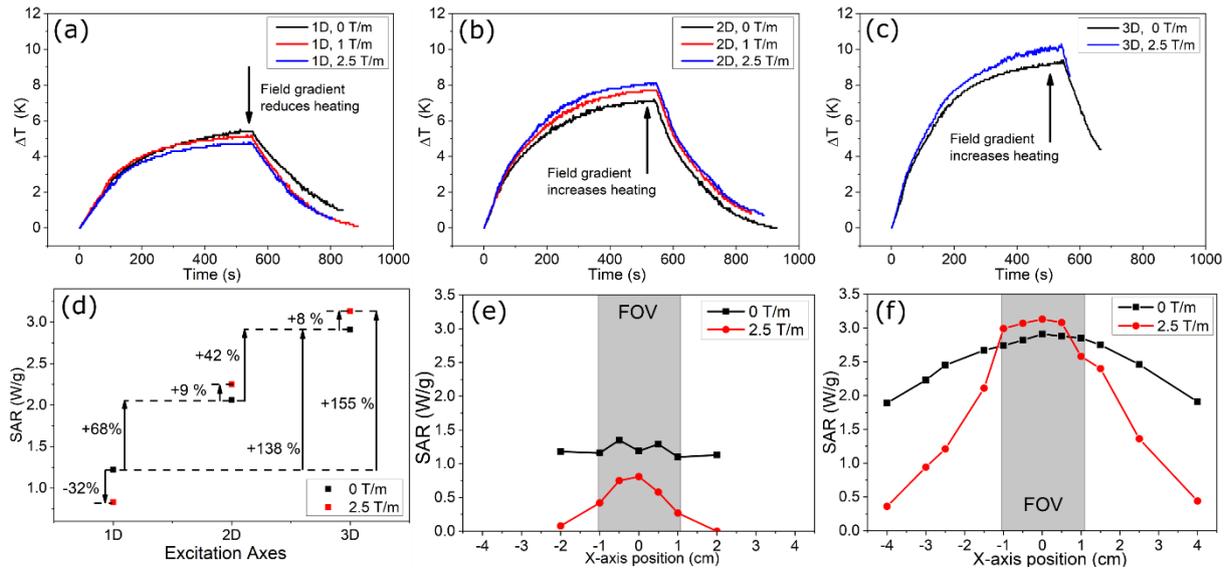

*Figure 2. Heating curves and influence of gradient fields for 50 μL FCT sample under (a)1D (b) 2D and (c) 3D excitations and varying static fields. d) plot showing SAR values calculated from experimental data for 1D, 2D and 3D excitations with zero and maximum field gradient. (e)-(f) x-axis spatial dependence of the SAR for (e) 1D and (f) 3D excitations both with, and without gradient fields.*

*Torque-based modeling of FCT heating*

Modelling results for FCT under MPI excitations are presented in Figure 3. In Figure 3 (a) the calculated Néel, Brownian and effective relaxation times for different core sizes of FCT are plotted, using constant values $R_0$ = 10$^{-9}$ s [31], K = 9 kJ/m³ [25] [32] , $k_B$ = 1.38 x10$^{-23}$ , $T$ = 293 K, $V_h$ = 60 nm [33], $\eta$ = 0.0010016 Pa.s [34]. Core sizes of 5.5 nm and 24 nm correlating to the two peaks of the bimodal size distribution are labelled. The results indicate that the smaller core size population within the FCT nanoparticles will be dominated by Néel relaxation, while the larger 24 nm cores will be Brownian dominated. This suggests that the use of the simulated torque calculation may be valid to describe the larger core distribution.

Figure 3(b) contains a table illustrating the calculated values for $|\overline{B_T}|$ and $\bar{\tau}$ obtained for (1D, 0 T/m), (2D, 0 T/m) and (3D, 0 T/m) excitations within the MPI scanner over a 44 ms period (time step = 1 μs, m = 1). Due to the lack of field gradients, these values are regarded as being homogeneous across the scanner's FOV. The results show that the inclusion of each additional excitation axis is associated with a significant, but non-uniform (+50% and +25%), increase in the average magnetic field vector length. Similarly, the torque value is seen to increase by 45% between the 2D and 3D cases, although a comparison cannot be made with the 1D case as the parameter cannot be calculated because the angles changes here only in two points in each cycle, during which the field amplitude is extremely small. The effect of adding a static gradient field in addition to the dynamic components is modelled for 3D and 2D excitations in Fig. 3(c-d), in both cases, adding the gradient field produces a strong focusing effect on the SAR, similar to that experimentally observed in Fig. 2(f). The 2D result showed evidence for an enhanced SAR within 1 mm of the center of the FOV, although this was not observed in the 3D case. The (3D, 2.5 T/m) modelling result at the center of the FOV was fitted to the experimentally measured SARs along this line in Fig. 3(e), showing some similarities. Using the maximum temperature increase observed in Fig. 1(b) for a 400 µL FCT sample located at the center of the scanner for calibration, a 2D map of the projected temperature increases for a similar sample at locations in the x- y- plane is presented in Fig. 3(f) based on the torque model.

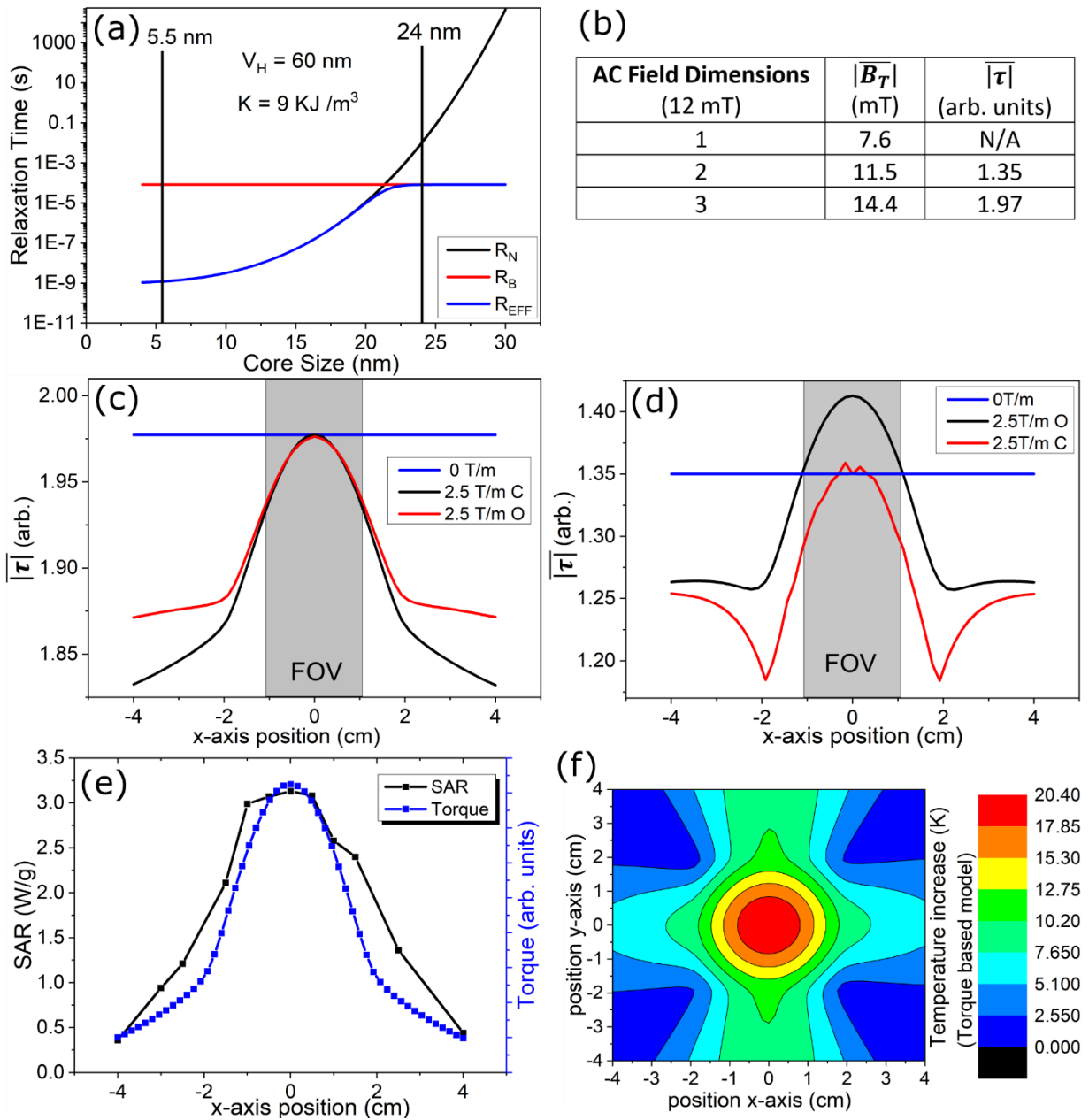

Figure 3. Modelling of FCT relaxation, MPI scanner's magnetic fields and exerted torques. (a) calculated Néel, Brownian and effective relaxation times for FCT nanoparticles with different core sizes. (b) $|\overline{B_T}|$ and $\overline{|\tau|}$ calculated for 1D, 2D and 3D excitations (12 mT) with no field gradient. (c-d) calculated $\overline{|\tau|}$ in the presence and absence of a gradient field, at positions along the scanner's x axis for 3D (c) and 2d, (d) excitations. "C" passes through FOV, "O" is offset 1 mm away from central axis. (e) Measured SAR data (Fig. 2(f)) fitted with theoretical curve based on $\overline{|\tau|}$. (f) 2D map of predicted heating at different locations in the scanner based on $\overline{|\tau|}$ modelling and experimentally observed temperature increase for 400 µL FCT sample in Fig. 1(b).

*Sample localization, hyperthermia application and real-time thermometry of tumor phantoms using MPI*

MPI results obtained during protracted exposure of a THP-1/FCT sample to a (3D, 2.5 T/m) imaging sequence are presented in figure 4. A photograph of the sample mounted on warm water pipes to maintain a 37 ˚C baseline temperature is shown in Fig. 4(a). A reconstructed 3D image of the MNP distribution is shown in Fig. 4(b), illustrating the ability of the scanner to reconstruct the location of MNPs distributed within THP-1 tumor phantoms. In Fig. 4(c) the changes in both the temperature of the sample (fiber-optic thermometer), and the $3^{rd}$ harmonic component of the raw MPI imaging signal ($A_3$) [35], are shown during a 40 minute exposure to a (3D, 2.5 T/m) MPI sequence within the scanner. A maximum temperature of more than 45 °C was recorded at the end of the treatment, placing the cells well above the temperature limit associated with MFH treatment. Amplitude of $A_3$ frequency component is plotted directly against the thermometer temperature in Fig. 4(d), showing an approximately linear correlation (Pearson's R = -0.978). To further demonstrate that the temperature sensitivity of the raw MPI Imaging signal translates into reconstructed images, images acquired at different temperatures and reconstructed using the same room temperature SF are presented in Fig. 4(e) using the "difference image" technique described in the materials and methods section. It is important to note that these are not conventional MPI images, but rather a demonstration of the changes induced in the reconstructions as a result of the changing sample temperature. The implication is that with access to SF measurements at different temperatures, calibrated images of the temperature at different locations within the scanner could be achieved. The difference images are taken at the timepoints labelled 1-6 in Fig. 4(c), A large change in the reconstructed intensity is registered between images 5(e) 1-4, which corresponds with the period in which both the sample temperature and $A_3$ amplitude are strongly changing in Fig. 4(c). Between images 5(e) 4-6 there is little change, which again corresponds well with the plateau in temperature and $A_3$ seen in Fig. 4(c) for these timepoints.

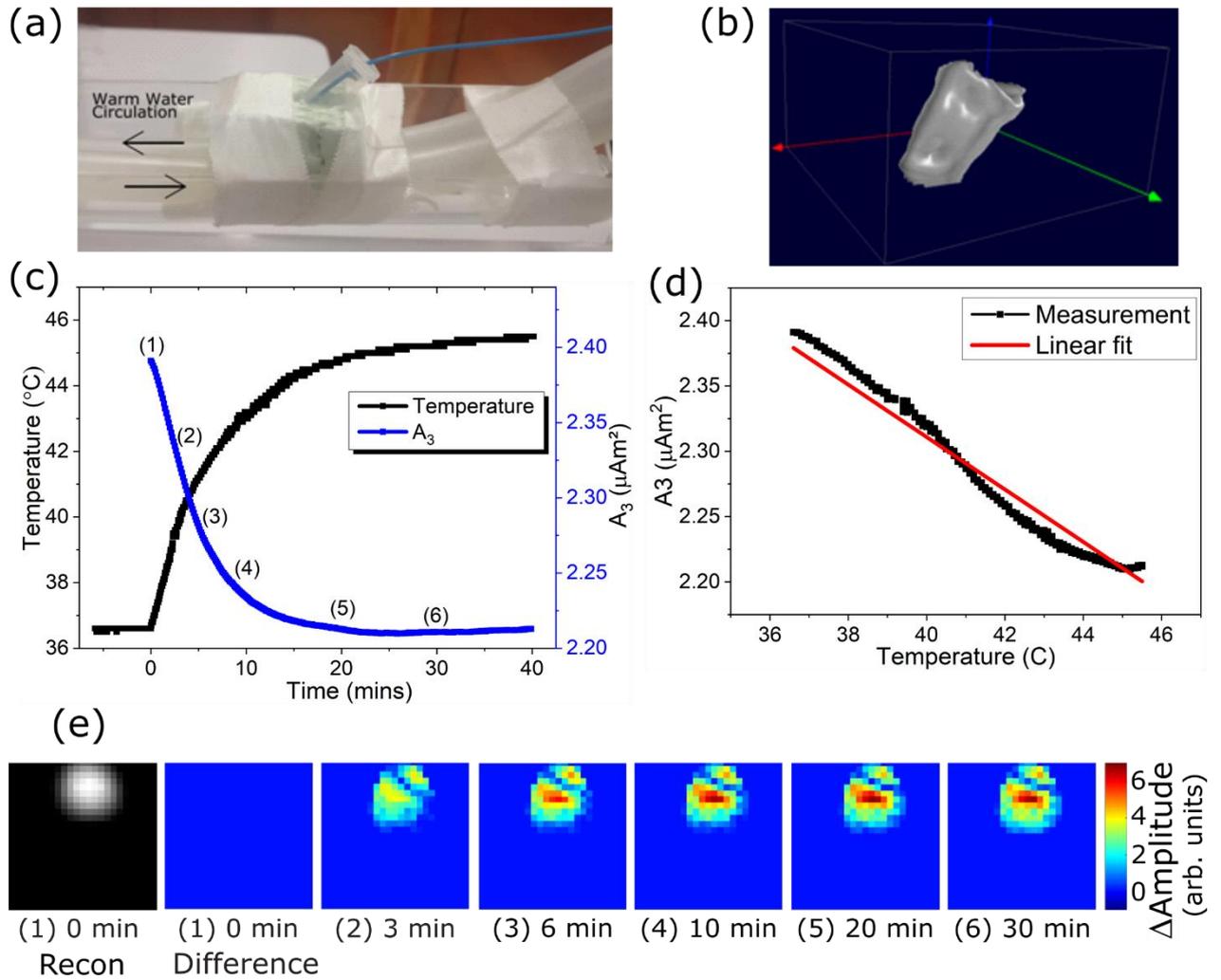

*Figure 4. (a) Photograph of THP-1/FCT sample mounted on warm water pipes for temperature control, ready for insertion into the scanner. (b) Reconstructed MPI image of phantom shown in (a). (c) Evolution of FCT-cell sample temperature (fiber-optic probe) and 3$^{rd}$ harmonic MPI signal amplitude $A_3$ during 40 min MPI exposure. (d) Temperature and signal amplitude data from (c) plotted against each other with linear fit. (e) difference images illustrating changes in image reconstruction as the phantom warms up. The first greyscale image shows the reconstructed nanoparticle distribution from above the sample at timepoint 1 in Fig. 4(c). The colorscale images show the difference images obtained for the reconstructions at timepoints labelled 1-6 in Fig. 4(c).*

*Cell viability following MPI-induced hyperthermia*

Next, the effect of MPI exposure on the cellular viability of THP-1 cells within loose tumor phantoms was investigated. The number of surviving cells 24 h after MPI exposure was estimated using the MTT assay. For all samples, the MPI-exposed samples showed reduced MTT absorbance compared with the control samples (Fig. 5a). These reductions indicated an average decrease in the number of viable cells within the MPI-exposed phantoms to 33.1 ± 11.5 % of the number observed in the control phantoms. To elucidate the underlying mechanism of cell death induced by MPI-exposure, flow cytometry analysis of membrane integrity and DNA distribution was performed. An example of the results of live cell DAPI staining for analysis of membrane integrity at time intervals during the 24 hours following MPI exposure are shown in Fig. 5b (upper panel), where increased DAPI fluorescence marks cells with lost membrane integrity. At 1 h after MPI exposure, the DAPI fluorescence was only minimally higher (4%) compared to control. In the final measurement 24 hours after treatment, almost half of the MPI exposed cells was recorded as DAPI-positive. The changes observed in Fig. 5b were broadly reciprocated in all experiments (n = 4). A summary of the average percentages of cells testing DAPI-positive at time points 1 h, 2 h, 4 h and 24 h after treatment in both the MPI exposed and control samples is shown in the lower panel of Fig. 5b. After 24 h the percentage of DAPI positives in MPI exposed samples had reached 50.2 ± 11.3 %.

For analysis of DNA distribution, DAPI stained samples were completely permeabilized with triton and measured by flow cytometry. The DNA distribution, i.e. the proportion of cells in the G1, S and G2 phases of the cell cycle, was similar in the MPI exposed and control samples at 24 h after treatment. No significant sub-G1 peak was observed in any sample, reflecting the absence of DNA fragmentation and thus the absence of a hallmark of apoptosis (Fig. 5c).

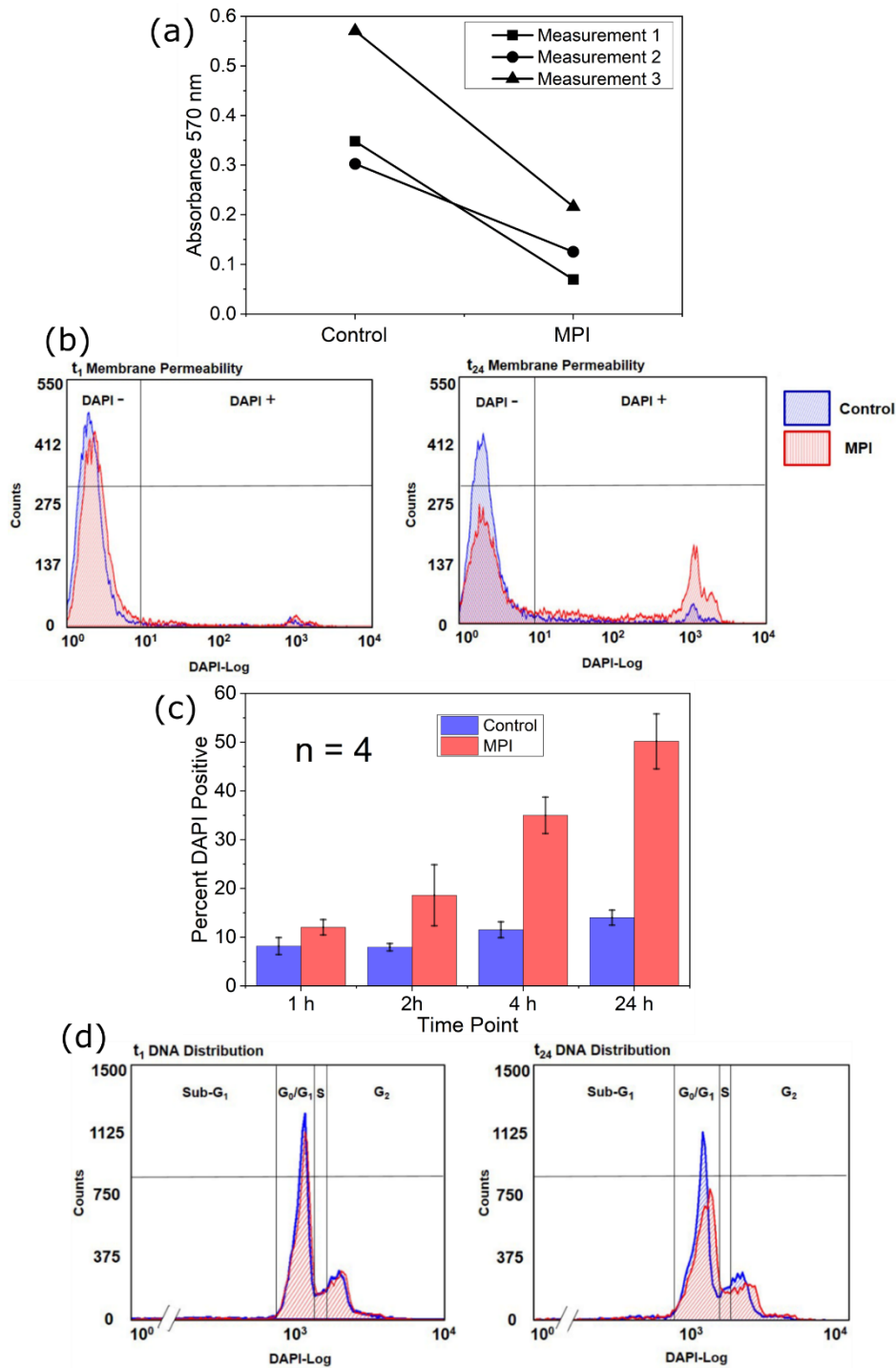

*Figure 5 (a) MTT assay results comparing relative absorbances at 570 nm of control and MPI exposed samples 24 hours after treatment. (b) Distribution curves obtained by flow cytometry represent the fluorescence intensity of DAPI stained MPI treated (red) and control (blue) cells at 1 h (left panel), and 24 h (right panel) after treatment. In each, the x-axis depicts the DAPI fluorescence intensity in logarithmic scale and the y-axis shows the cell count. (c) Percentage of DAPI positive THP-1 cells in MPI-exposed and control samples 1 h, 2 h, 4 h and 24 h after MPI exposure (Data are presented as means with SD, n=4 experiments). 5 (d) enlarged DNA distribution curves from flow cytometry results in (b). G1 peak = diploid chromosome content, G2 peak = doubled diploid chromosome content, S=intermediate chromosome content in S phase. Note the absence of subG1 peak representing DNA fragments (hallmark of apoptotic cell death).*

## Discussion

Calorimetry measurements within the 20/25 FF MPI scanner using 400 µL FCT ($c$(Fe)=935 mmol/L) samples revealed non-negligible and reproducible MNP hyperthermia. The extent of this heating was heavily influenced by both the number of excitation axes implemented, and the presence or absence of static field gradients. Maximum heating was observed for the excitation sequence associated with the highest resolution 3D image acquisition sequence (3D, 12 mT). This imaging sequence was shown to heat the FCT sample by around 5 K over the first 2.5 minutes of field exposure. By contrast, negligible heating was previously reported during 2.5 minute exposure times when 100 µL samples were subjected to continuous imaging by a raster scanning field-free-line based MPI scanner using a 1D excitation ($f$ = 20 kHz, $H$ = 20 mT) and 2.35 T/m field gradient [21]. There are likely to be many reasons contributing to these differing results. The 25 % larger frequency within the 20/25 FF Lissajous scanner will play a role. The concentration of nanoparticles used in [21] was also significantly less than was used in the study reported here. Furthermore, different types of nanoparticles were used in each study, which may themselves have different heating efficiencies. It is also possible that the raster scanning trajectory used in [21] prevents heat from accumulating in the sample to the same extent as in Lissajous scanning. However, with the currently available data it is not possible to empirically separate the extent of each of these potential contributions. A more detailed collaborative study would be required in the future to develop a detailed understanding of the differences in the heating observed in each scanner type.

The maximum recorded temperature rise recorded in the Lissajous scanner was observed for a (3D, 2.5 T/m) excitation (Fig. 1). This magnetic field sequence is commonly employed for high-resolution preclinical image acquisition using the device. The 9 minute exposure time used for calorimetry measurements is long compared with a single imaging cycle in this system, although similar acquisition times have been reported in other MPI studies [8]. It is likely that the larger SARs observed when the excitation field increases from 1D to 2D, or 2D to 3D, are partially related to the increasing values of $|\overline{B_T}|$ as shown in Figure 3. Typically, the SAR is expected to increase linearly with the frequency, and quadratically with increasing field amplitude [36]. The reduction and spatial focusing of the SAR observed when a static gradient field is added to a 1D excitation (Fig. 2(e)) is consistent with results previously reported [37]. The focusing effect of the SAR under (3D, 2.5 T/m) conditions is less narrow than in the (1D, 2.5T/m) case, although still present. The precise cause of this broadening in the 3D case is not completely understood at present and will be the subject of additional modelling and experiments. The presence of spatially focused heating within the Lissajous MPI scanner suggests that the system has the potential to be used for tumor-selective focused MFH treatment, similar to previously described in [19] [21].

SAR enhancement is observed experimentally at the center of the scanner when a static field gradient is added to a 2D or 3D excitation (Figs 1 and 2). This enhancement is believed to be a real effect as the measurements with static field on and off were conducted consecutively at each location, before moving the sample to the next position. This process minimized the possibility of a miscalibration between the results. The possibility of additional heat leaking from the scanner electronics into the sample was dismissed via the measurement of a pure water sample under maximum fields, with no temperature change. Some aspects of the experimentally observed SARs, and their spatial dependences, have been qualitatively described by the modelling of $|\overline{B_T}|$ and $\overline{|\tau|}$ as presented in Fig. 3. $\overline{|\tau|}$ is necessary to describe the spatial distribution of the SAR in the presence of a field gradient, as $|\overline{B_T}|$ shows only an increase in amplitude with increasing distance from the FFP. The broader peak in the experimental SAR vs the model in Fig. 3(e) is probably related to smearing created by the non-zero volume of the FCT sample and heat diffusion effects. The calculated relaxation times (Fig. 3(a)) suggest that the torque model is applicable to the larger core population present within FCT, but not the smaller. Recent experimental studies have indicated that the majority of the MPI signal and heat generated by FCT originate from this population of larger aggregates [38] [39]. The full extent of the mechanisms underlying the heating behavior observed within the scanner is expected to be more complex than the torque-based model, and lies beyond the scope of the present study. The torque-based model is expected to only be applicable in the lower-frequency, Brownian dominated, regime. Additional measurements spanning multiple frequencies could explore this.

MPI imaging results (Figure 4) demonstrate different capabilities of the scanner which can complement MFH therapy. The scanner is shown to be capable of reconstructing accurate 3D images of MNP distributed among cancer cells, suitable for MNP localization within a tumor (Fig. 4). The change in $A_3$ as the THP-1/FCT sample heats up demonstrates the temperature sensitivity of the raw MPI imaging signal, and the potential for MPI-based remote thermometry during hyperthermia treatment. This effect is due to the thermal sensitivity of both Néel and Brownian relaxation processes [28], and has been previously demonstrated as the basis for temperature resolved 3D MPI images [22]. Slight deviations from linear dependence are observed in Fig. 4(d), this is attributed to additional influences on the MNP dynamics such as gradual loss of mobilization during MNP-cell contact and/or internalization. Similar cellular interaction effects have been previously studied using different nanoparticle/cell combinations in the same scanner [40]. While these secondary effects could increase the uncertainty of temperature measurements using this technique, it is likely that techniques for minimization or compensation can be developed. MPI image reconstructions for the THP1/FCT sample at different temperatures also show changes, highlighted in the difference images shown in Fig. 4(e). This further emphasizes that with suitable infrastructure for SF acquisition at different temperatures, the scanner can be used for 3D imaging of the tumor temperature during MPI based MFH therapy. The development of an apparatus suitable SF acquisition at different temperatures is now underway. Once completed, a multicolor reconstruction technique using two SFs measured at differing temperatures can be used to reconstruct temperature maps using the same reconstruction concepts detailed in [40].

Despite the relatively low frequencies employed in Lissajous scanning MPI, prolonged imaging exposures induced hyperthermia conditions within a tumor phantom formed of THP-1 monocytic leukemia cells and FCT, resulting in significant cell death. Tumor cells, including THP-1, are expected to have a high thermal tolerance due to a high expression of heat shock proteins [41]. The MPI-induced loss of more than 60 % of the cell count at 24 h after exposure is therefore remarkably high. Previous studies showed that monocytic THP-1 cells do not efficiently internalize FCT [42] , thus it is assumed that the MPI-induced heat occurred predominantly in the local environment of THP-1 cells. Massive cell death did not occur directly during MPI exposure but rather over the subsequent 24 h period following exposure. A delayed onset in the loss of cellular viability, at a temperature similar to that measured after MPI induced heat using FCT (45°C), was also reported for phantoms of MNP loaded neuroblastoma cells post magnetic hyperthermia [43]. The authors suggested that surface heating due to the presence of MNPs on the cell surface caused crater-like pores in the cell membrane, which were detected using electron microscopy. This eventually led to cell death through a combination of necrosis and apoptosis. We propose that a similar phenomenon may be responsible for the results reported in our study. The stepwise increase in membrane permeability, measured by DAPI uptake, points to a stepwise disruption of the cell membrane as the trigger of cell death after MPI induced hyperthermia, using FCT. The lack of DNA fragmentation in dying THP-1 cells excludes apoptosis and rather points to necrosis as the underlying mechanism of cell death. Clearly, not all cells of the tumor phantom died after MPI exposure. Reasons for this can include a higher thermal tolerance of one subpopulation of cells within the sample, or an inhomogeneous heat distribution within the phantom. The images in Fig. 4(e) may indicate higher temperatures at the center of the phantom as compared with the edges, although further study is required to confirm this.

The cell studies reported here employed leukemia monocytes in loose tumor phantoms. Further work would be required to ascertain whether these results obtained under in-vitro conditions translate into real tissue damage in-vivo. Key considerations which could reduce the potential for damage in-vivo include the inhibited dynamics of MNP when lodged within tissue, and the heat-transport effect that blood-flow through organs may exert. This study has only dealt with a single type of MNP tracer, and the field conditions experienced in a specific MPI scanner type. Additional work is required to identify the wider effects produced by other MNP types and scanner designs. It is possible that the field conditions within other types of scanner may result in significantly different heating behaviors.

The application of a therapeutic MFH dose using an MPI scanner's imaging excitation has not been reported before. This technique offers possible advantages compared to the dual excitation MFH-MPI technique reported in [21], including the potential for continuous 3D thermal imaging of the targeted tissue during therapy. Using Lissajous scanning MPI, the location of the hyperthermia agent within a tumor can first be verified using a short imaging sequence. Following this, simultaneous application of focused therapeutic heating and remote real-time temperature monitoring can be realized using a second prolonged exposure to the same magnetic field sequence. Used in this way, the authors believe that Lissajous scanning MPI can greatly enhance the precision and effectiveness of MFH therapy. It should however be noted that the heat transfer processes occurring in-vivo may differ significantly from those in the in-vitro measurements reported here. Blood flow can remove a significant amount of heat from the site of hyperthermia, reducing the overall temperature rise. Conversely, we have focused on relatively small volumes of tracer in this work. If in-vivo applications employ larger volumes of MNP, the surface area to volume ratio may change sufficiently to reduce the rate of heat loss, allowing higher temperatures to be reached. Additional work is needed to explore these factors and establish the effects of Lissajous MPI actuated hyperthermia in-vivo.

With respect to pure MPI development, this study demonstrates that there remains a need for more in-depth analysis of the impact of MPI on tissues. It is necessary to ascertain the safe limits for imaging protocols for a given type of scanner before the technique can be verified as safe by regulatory bodies. It should be noted that the MNP concentrations and field exposure times implemented in this work are much greater than the minimums necessary for conducting basic Lissajous MPI imaging. It is likely that a significant window exists at low MNP concentrations and field exposures in which Lissajous MPI does not inflict damage. Further work to establish in detail the safe limits for nanoparticle concentrations and field exposures during Lissajous MPI will be undertaken in the near future. Of course, the work presented here has focused on the impact of heating effects produced in-vitro by a single type of MPI scanner and a single nanoparticle system. It is possible that other approaches to the design of MPI scanner architectures, field trajectories and imaging sequences may exhibit less heating than reported here. For example, alternative scanning techniques such as those based on raster scanning or the travelling wave technique may result in lower SAR values [9] [21]. Other types of nanoparticle may also exhibit radically different heating behaviors compared to the FCT employed here. Like the spatial resolution of MPI, the SAR focusing effect provided by the scanner's gradient field will depend strongly upon the M-H curve of the nanoparticles used. Improved spatial focusing of MFH treatment and other optimizations may be achieved via the exploration of alternative nanoparticle types.

## Conclusion

The heat dissipated by FCT tracer during Lissajous scanning MPI has been studied via calorimetric measurements complemented by a simple torque-based model. The highest energy dissipation was registered at the center of the scanner's FOV during excitation with magnetic field sequences typical for high-resolution 3D image acquisition. We were able to show that the heat dissipation generated in this way induced cell death in loose tumor phantoms, due to a thermally induced partial cell membrane damage, which subsequently led to destruction of around 50 % of cell membranes within the 24 hours following treatment.

The evidence presented here shows that heating effects are an important consideration in MPI development, offering both a source for caution, and potential opportunities for high-impact applications. In particular, the ability of Lissajous scanning MPI to act as a multi-faceted platform to aid in the application of MFH cancer therapy has been highlighted here for the first time. The technique offers a platform in which the location of the heating agent can be verified in-vivo, a tumor-specific focused excitation field can be applied to actuate the therapy and the temperature of the tumor tissue can be remotely monitored during treatment. Furthermore, the results presented here emphasize that the nanoparticle concentration administered to patients, and the destination and density of those nanoparticles within the body should be considered carefully before MPI data acquisition is undertaken. Protracted MPI exposures for the monitoring of dynamic processes, or the averaging of weak signals, should only be undertaken in-vivo with due care and consideration.

## Conflicts of interest

There are no conflicts to declare.

## Acknowledgements

*This project was supported by the DFG research grants "quantMPI: Establishment of quantitative Magnetic Particle Imaging (MPI) application oriented phantoms for preclinical investigations" (grant TR 408/9-1) and "Matrix in Vision" (SFB 1340/1 2018, no 372486779, projects A02 and B02).*

# Appendix

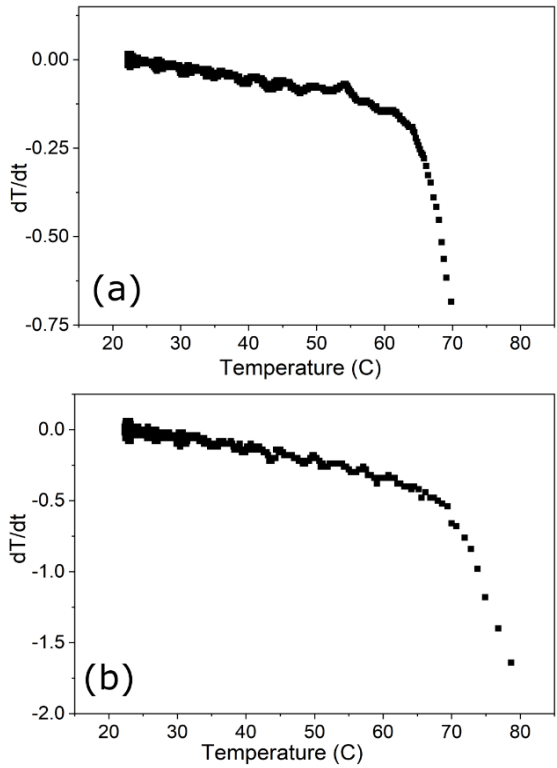

*Appendix 1 – Numerical derivatives of cooling curves obtained for the (a) 400 µL FCT and (b) 50 µL FCT samples mounted in the MPI scanner. The results identify the extent of the linear-loss regime for each sample size and holder, and verify the use of the corrected slope technique for SAR analysis of the heating curves obtained from each sample volume within the experimental set-up.*